%% file: Dh312_astroph.tex
\documentclass{aa}
\usepackage{aabib}
\usepackage{graphicx}
\input{definition_journaux_aasms4.tex}

\begin{document}
   \title{Discovery of new embedded Herbig-Haro objects\\
in the $\rho$ Ophiuchi dark cloud\thanks{Based on observations carried out 
at the European Southern Observatory, La Silla, Chile, under project 67.C-0325(A).}}

\titlerunning{New embedded Herbig-Haro objects in the $\rho$ Ophiuchi dark cloud}
\authorrunning{Grosso et al.}


   \author{N. Grosso\inst{1}
      \and J. Alves\inst{2}
      \and R. Neuh{\"a}user\inst{1}
      \and T. Montmerle\inst{3}
 	  }

   \offprints{N. Grosso, ngrosso@mpe.mpg.de\,.}

   \institute{Max-Planck-Institut f{\"u}r extraterrestrische Physik,
      	      P.O. Box 1312, D-85741 Garching bei M{\"u}nchen, Germany
	 \and European Southern Observatory
	      Karl-Schwarzschild-Str. 2,
       	      D-85748 Garching bei M{\"u}nchen, Germany
	 \and Service d'Astrophysique, 
     	      CEA Saclay, 
	      F-91191 Gif-sur-Yvette, France \\
             }

   \date{Received ; accepted }

   \abstract{We report here the discovery of a 30\arcsec-chain of embedded Herbig-Haro (HH) objects in the $\rho$ Ophiuchi dark cloud. These HH objects were first detected during a deep $K_\mathrm{S}$-band observation (completeness magnitude for point source$\sim$19) made with NTT/SOFI. We confirm their nature with follow-up observations made with H$_2$ {\it v}=1--0 S(1) narrow-band filter. 
We argue that they belong to two different jets emanating from two Class~I protostars: the main component of the recently resolved subarcsecond radio binary \object{YLW15} (also called \object{IRS43}), and \object{IRS54}. We propose also to identify the [S\,{\footnotesize II}] knot \object{HH224NW1} (G{\' o}mez et al.\ 1998) as emanating from a counterjet of YLW15. The alignment between these HH objects and the thermal jet candidate found in YLW15 by Girart et al.\ (2000) implies that this jet is not precessing at least on timescale $\sim$(2--4)$\times$10$^4$\,yr. 
               }

   \maketitle
%

\section{Introduction}

In the 1950's, objective prism surveys of dark clouds revealed optical 
small nebulae, associated with young stars, showing emission line spectra 
with very weak continua. Today, these Herbig-Haro (HH) objects are interpreted 
as shocks produced by the interaction of outflows from Young Stellar 
Objects (YSOs) and the interstellar medium (see review by \cite{reipurth99}). 

The $\rho$ Ophiuchi dark cloud is one of the nearest (d$\sim$140\,pc) 
active site of low-mass star formation, displaying a rich embedded 
cluster of $\sim$200 YSOs (see the updated census by the ISOCAM 
survey; \cite{bontemps01}). It is thus one of the most suitable locations 
for the search of HH objects. Optical emission-line surveys 
(H$\alpha$, [S\,{\small II}]) have led to the detection of only 10 bona 
fide HH objects (\cite{reipurth88}; \cite{wilking97}; \cite{gomez98}; 
\cite{reipurth_cat99}). As outlined by Wilking et al.\ 
and G{\' o}mez et al., most of these HH objects are located at the periphery 
of the cloud, in the lowest extinction area: optical surveys do not 
have access to embedded HH objects. \cite*{wilking97} noted for all these 
nebulae a relatively high [S\,{\small II}]/H$\alpha$ ratio, which is a tracer 
of low excitation conditions, and thus indicates that H$_2$ is not 
destroyed after the bow shock; they suggested to use 
the shocked-H$_2$ transition at $\lambda$=2.121\,$\mu$m to probe deeply 
embedded HH objects.
Up to now in the $\rho$ Ophiuchi dark cloud, shocked-H$_2$ imaging was only performed  
to study molecular shocks in the CO outflow of the Class~0 protostar prototype \object{VLA1623} 
(e.g., \cite{davis95}), which led to the detection of several H$_2$-knots; 
one of them was later detected in [S\,{\small II}] (\object{HH313}; \cite{gomez98}). 

We report here deep near-IR observations of the $\rho$ Ophiuchi dark cloud
unveiling bow-shape structures and knots, and complementary observation 
with narrow-band filter confirming their nature as H$_2$ shocks. 
As these objects are not optically visible, we called them 
{\it embedded} HH objects, and discuss the possible exciting sources.

\section{Observations and results}

\begin{figure}[t]
   \centering
   \includegraphics[width=0.8\columnwidth]{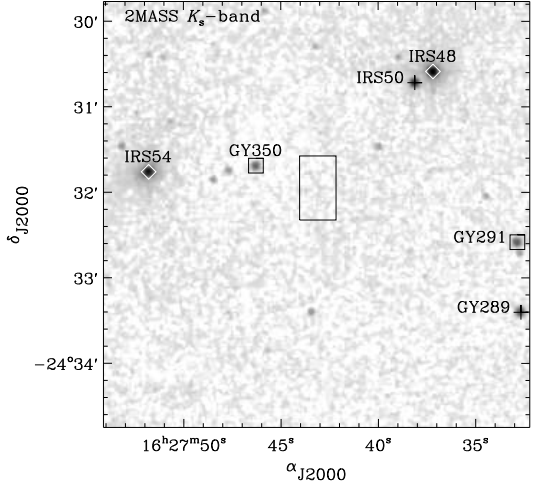}
      \caption{{\footnotesize $K_\mathrm{S}$-band finding chart based on 2MASS survey image,  
showing the NTT/SOFI pointing field of view. 
The YSOs of this area are marked according to their IR classifications 
(diamond/square/cross for {\it resp.} Class~I/II/III IR source; see Lada 1991) 
from the ISOCAM mid-IR survey (Bontemps et al.\ 2001).   
The box indicates the area enlarged in Fig.~\ref{hh}a.} 
              }
\vspace{-0.4cm}
         \label{finding_chart}
   \end{figure}

During the period April 4--7 2001, deep $J, H, K_\mathrm{S}$-band observations 
of the $\rho$ Ophiuchi dark cloud were made by N.G. with NTT/SOFI, as follow-up 
of Chandra and XMM-Newton X-ray observations.
The complete report of this follow-up, consisting of 5 pointings, 
will be published in an upcoming paper (Grosso et al., in preparation). 
We will focus here only on the pointing related to the detection of 
new embedded HH objects, and located SE of the millimetric 
dense core Oph-B2 (\cite{motte98}; see Fig.~\ref{finding_chart}). 
This area was already surveyed by optical emission-line surveys, 
but no HH objects were found.

We used the auto jitter imaging mode to obtain a total integration time of 40\,mn, 
with 6\,s detector integration time (DIT) and 1\,mn-exposure frames.
The near-IR standard AS29-1 (\cite{hunt98}) was observed several times during the night.
We took five 6\,s-darks at the end of the night.  
The different stages of the usual IR data reduction, namely sky estimation and subtraction, 
frame recentering and stacking, including also dark substraction and flat-field division 
(we used the flat-field provided by the NTT team for our observational period), 
were performed using the {\tt jitter} routine of the ESO's {\tt eclipse} package 
(version 3.8-1). 
Zero order astrometric correction was applied using the 2MASS field stars 
(\cite{cutri00}) as reference frame, reducing the offset residual to 0.7\arcsec.
The magnitude zero-point was obtained from our calibration star.

\begin{figure*}[t]
   \centering
   \includegraphics[angle=90,width=\textwidth]{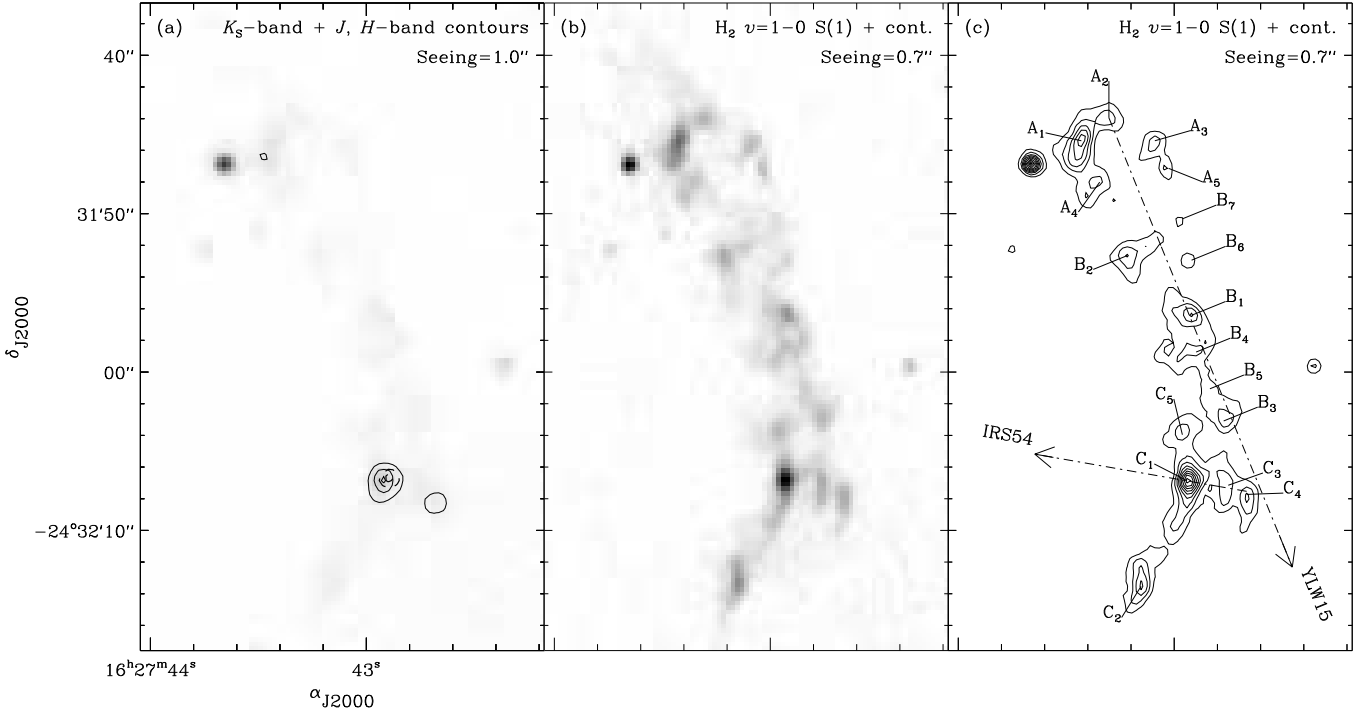}
      \caption{{\footnotesize Deep near-IR images of the embedded HH objects discovered with NTT/SOFI.
(a) $K_\mathrm{S}$-band image scaled so that continuum features from stars 
appear as bright as in the H$_2$ {\it v}=1--0 S(1) narrow-band filter image.
The pixel size is 0\farcs29, and the grey color scale is linear.
$J$-band ({\it resp.} $H$-band) detections above 5$\sigma$ levels are shown with dashed 
({\it resp.} continuous) contour levels (24.5, 23.5, 23.0 and 21.8, 21.2, 
20.9\,mag\,arcsec$^{-2}$).
(b) H$_2$ {\it v}=1--0 S(1) narrow-band filter image (including the continuum emission). 
H$_2$ line emission features appear brighter in this narrow-band image than in the previous 
broad-band one.
(c) H$_2$ {\it v}=1--0 S(1) contour map (18.3, 17.6, 17.2, 16.9, 16.6, 16.5, 16.3, 16.1, 16.0, 15.9 \,mag\,arcsec$^{-2}$). 
The H$2$-knots are labeled by decreasing H$_2$ magnitudes (see Table~\ref{tab:knot}).
The first ({\it resp.} second) arrow joins the barycenter of the knots A--B ({\it resp.} C) 
and the Class~I protostar YLW15 ({\it resp.} IRS54).
The first arrow corresponds also to the symmetry axis of the knots A--B.}}
\vspace{-0.4cm}
         \label{hh}
   \end{figure*}

Fig.~\ref{hh}a is an enlargement of the observation obtained with the $K_\mathrm{S}$ 
broad-band filter ($\lambda$=2.162\,$\mu$m, $\Delta\lambda$=0.275\,$\mu$m), 
unveiling a complex object showing bow-shape structures and possible emission knots.
To detect weak structures in the $J$ and $H$-band images, 
we filtered it with the wavelet transform ({\tt MR/1} package; \cite{starck98}) 
to a 5$\sigma$ significance threshold.
In this manner, a few of knot candidates are also detected in the $H$-band, 
and only one is detected in the $J$-band.
The shape of this complex object is reminiscent of the ones found in HH objects. 
If this HH object interpretation is correct, this $K_\mathrm{S}$-band detection 
must be mainly due to the 2.121\,$\mu$m-line of the H$_2$ shocked emission. 
To confirm it, J.~A.\ made a 20\,mn-exposure observation (DIT=30\,s, 1\,mn-exposure frames) 
with NTT/SOFI using the H$_2$ {\it v}=1--0 S(1) narrow-band filter ($\lambda$=2.124\,$\mu$m, 
$\Delta\lambda$=0.028\,$\mu$m) during the night of June 8 2001 (Fig.~\ref{hh}.b). 
The flat-field was computed from the median of the dark substracted frames.
The good seeing conditions (FWHM$\sim$0.7\arcsec) of this shorter observation gives 
much more details than the $K_\mathrm{S}$-band image.  

To differentiate between pure line emission knots and continuum emission from stars 
(including scattered light) we follow \cite*{davis95} who noted that the narrow-band filter 
reduces the intensity of the field stars by the ratio of the filter bandpasses ($\sim$10),
whereas the intensity of the H$_2$ knots is only reduced by $\sim$2 
(indeed spectra of H$_2$ molecular shocks have other emission lines in the $K_\mathrm{S}$-band; 
see \cite{smith95}). We scale down the $K_\mathrm{S}$-band image by the ratio of the filter 
bandpasses and the ratio of the seeing to have continuum features appearing with the same 
brightness in Fig.~\ref{hh}a and \ref{hh}b. 
On the other hand, H$_2$-shocks appear $\sim$5 times brighter 
in Fig.~\ref{hh}b than in Fig.~\ref{hh}a and can thus be easily identified.
The H$_2$-knots detected are labelled on the H$_2$ contour map (Fig.~\ref{hh}c).
Table~\ref{tab:knot} gives the position of these knots  
with photometry for a 1\arcsec-aperture from the 5$\sigma$-filtered images.
Before applying wavelet filtering the H$_2$ image was convolved with a Gaussian filter 
to have the same seeing as in the $K_\mathrm{S}$-band image. 
To obtain the H$_2$ magnitudes, we scaled up the H$_2$ intensities 
by the filter-band width ratio and applied the magnitude zero-point of 
the K$_\mathrm{S}$ image.

The knots A display a bow-shape structure, characteristic of low-speed 
bow-shocks which produce such arcs and limb-brightened structures of conical appearance 
(\cite{smith91}). 
We also note in the H$_2$ image some diffuse emission downstream of the bow-shock
which may be related to the excitation of H$_2$ in the pre-shock region of a J-shock. 
We thus propose the knots A as the leading part of a jet coming from the SW direction, 
and outlined by the upstream knots B.
By contrast the knots C are displaced from the direction of this jet, 
and moreover the knots C$_1$--C$_4$ are clearly elongated, roughly along a NS direction. 
We propose to explain these features by shocks coming from the East 
and producing the arc C$_2$-C$_1$-C$_5$, with the downstream knots C$_3$ and C$_4$.
We will discuss in the following section the possible exciting sources of these two jets.

\begin{table}[!hb]
\caption[]{H$_2$-knot positions and magnitudes.}
\label{tab:knot}
\resizebox*{\columnwidth}{!}{
\begin{tabular}{@{}cccccccc@{}}
\hline
\noalign{\smallskip}
Knot & $\alpha_{\mathrm{J2000}}$  &  $\delta_{\mathrm{J2000}}$ &  $J$ & $H$ & $K_\mathrm{S}$ & H$_2$ & f($K_\mathrm{S}$)  \\
names & 16$^{\mathrm h}$27$^{\mathrm m}$  &  -24$\degr$  &  [mag]  &  [mag] &  [mag]  &  [mag] & $\overline{\mathrm{f(H}_2\mathrm{)}}$  \\
\noalign{\smallskip}
\hline
\noalign{\smallskip}
  A$_1$ & 43$\fs$4 & 31$\arcmin$45$\farcs$4 &   -- & 23.7 & 19.3 & 17.1 & 1.4 \\
  A$_2$ & 43$\fs$3 & 31$\arcmin$44$\farcs$0 &   -- &   -- & 20.1 & 17.9 & 1.3 \\
  A$_3$ & 43$\fs$1 & 31$\arcmin$45$\farcs$4 &   -- &   -- & 20.3 & 18.1 & 1.2 \\
  A$_4$ & 43$\fs$0 & 31$\arcmin$47$\farcs$1 &   -- &   -- & 20.2 & 18.1 & 1.4 \\
  A$_5$ & 43$\fs$4 & 31$\arcmin$48$\farcs$0 &   -- &   -- & 20.9 & 18.6 & 1.1 \\
\noalign{\smallskip}
  B$_1$ & 42$\fs$9 & 31$\arcmin$56$\farcs$4 &   -- &   -- & 19.7 & 17.6 & 1.4 \\
  B$_2$ & 43$\fs$2 & 31$\arcmin$52$\farcs$7 &   -- &   -- & 20.0 & 17.7 & 1.2 \\
  B$_3$ & 42$\fs$8 & 32$\arcmin$03$\farcs$1 &   -- &   -- & 19.9 & 17.8 & 1.5 \\
  B$_4$ & 42$\fs$9 & 31$\arcmin$58$\farcs$7 &   -- &   -- & 20.0 & 17.9 & 1.4 \\
  B$_5$ & 42$\fs$8 & 32$\arcmin$01$\farcs$1 &   -- &   -- & 20.6 & 18.4 & 1.3 \\
  B$_6$ & 42$\fs$9 & 31$\arcmin$52$\farcs$9 &   -- &   -- & 21.8 & 18.8 & 0.6 \\
  B$_7$ & 43$\fs$0 & 31$\arcmin$50$\farcs$3 &   -- &   -- & 21.8 & 19.0 & 0.7 \\
\noalign{\smallskip}
  C$_1$ & 43$\fs$0 & 32$\arcmin$06$\farcs$9 & 24.5 & 21.2 & 18.5 & 16.6 & 1.7 \\
  C$_2$ & 43$\fs$2 & 32$\arcmin$13$\farcs$5 &   -- &   -- & 19.6 & 17.5 & 1.4 \\
  C$_3$ & 42$\fs$7 & 32$\arcmin$07$\farcs$7 &   -- &   -- & 19.6 & 17.8 & 2.0 \\
  C$_4$ & 42$\fs$8 & 32$\arcmin$07$\farcs$2 &   -- & 22.2 & 19.7 & 17.8 & 1.8 \\
  C$_5$ & 43$\fs$0 & 32$\arcmin$04$\farcs$0 &   -- &   -- & 20.0 & 17.9 & 1.4 \\
\noalign{\smallskip}
\hline
\noalign{\smallskip}
\end{tabular}}
\end{table}

To obtain an estimate of the H$_2$-shock velocity we compare the observed near-IR 
photometry with the predictions for planar molecular shocks (\cite{smith95}). 
We update first these results for the NTT set of filters. 
We extract the filter transmission profiles from the plots provided on the SOFI web 
page, and convolve them to the unfiltered molecular shock line fluxes (Michael D. Smith, 
private communication). Zero points for $J-H$ and $H-K_\mathrm{S}$ colors are derived 
by requiring the colors of A0 spectrum template (\cite{pickles98}) to be zero. 
We observe f($K_\mathrm{S}$)/f(H$_2$)$\sim$1.4 and $\sim$2.0 for the leading part 
of the two jets, which must be compared to f($K_\mathrm{S}$)/f(H$_2$)=2.0--3.3 
({\it resp.} 1.9--3.3) for J-shocks ({\it resp.} C-shocks) with velocity range 
8--22\,km\,s$^{-1}$ ({\it resp.} 20--45\,km\,s$^{-1}$). 
This implies J-shock ({\it resp.} C-shock) velocity $\le$10\,km\,s$^{-1}$ 
({\it resp.} 20\,km\,s$^{-1}$), well below the H$_2$ dissociation velocity 
in molecular cloud ($\sim$22, $\sim$47\,km\,s$^{-1}$ for {\it resp.} J, C-shock; 
see \cite{smith95}). 
Hence, the H$_2$-line emission is really mapping the bow-shock; 
we can exclude the scenario where this emission would come only from the low-velocity wing 
of a fast dissociative bow-shock located downstream, with strong iron lines in the $J$ and 
$H$ bands tracing the apex.
These low-velocity embedded H$_2$-shocks are reminiscent of the low excitation HH objects 
already observed in the optical in this dark cloud.  
From the intrinsic color of low-velocity H$_2$-shocks, $(J-H)_0$$\sim$0.8, 
and assuming the reddening law quoted by \cite*{cohen81}, we estimate the extinction 
for knot C$_1$: $A_\mathrm{V}$=9.09$\times$[($J-H$)-($J-H$)$_0$]$\sim$23.
This large visual extinction explains the non-detection by previous optical surveys.
It will be possible to measure directly $A_\mathrm{V}$ by $J$, $H$-band spectroscopy using 
[Fe\,{\small II}] lines at 1.25\,$\mu$m, 1.64\,$\mu$m, which arise from the same atomic upper level.

\section{Discussion on the exciting sources}

The optical survey of HH objects in Taurus has shown that their frequency 
decreases rapidly with the age of the YSO, from Class~I sources 
to Class~II sources (\cite{gomez97}), i.e. from evolved protostars 
to classical T~Tauri stars. 
This result is consistent with the decrease of the CO outflows 
observed from the Class~0 sources to the Class~I sources (\cite{bontemps96}).
To find the exciting source candidates of the HH objects reported here,
it is thus reasonable to look for YSOs with IR excesses.
On the basis of the HH feature morphology, we proposed in the previous section  
shocks coming from the East ({\it resp.} SW) to explain the shape of knots C 
({\it resp.} A--B).
We checked by constructing a color-color diagram of the sources detected 
in our deep observation, 
that there is no new embebbed YSO with IR excess in these directions.
Two known YSOs are Eastward (see Fig.~\ref{finding_chart}): 
the Class~II source \object{GY350}, and the Class~I protostar \object{IRS54}. 
We propose to associate knots C with IRS54, the agreement with the knots C shape 
looking better (see Fig.~\ref{hh}.c). 
Strong H$_2$ {\it v}=1--0 S(1) line emission was detected in the IRS54 spectrum 
(\cite{greene96}), it is unresolved in our H$_2$ image, which shows only scattered 
light Eastward.
To our knowledge this source has never been included in a CO outflow survey.

    \begin{figure}[!b]
   \centering
   \includegraphics[width=\columnwidth]{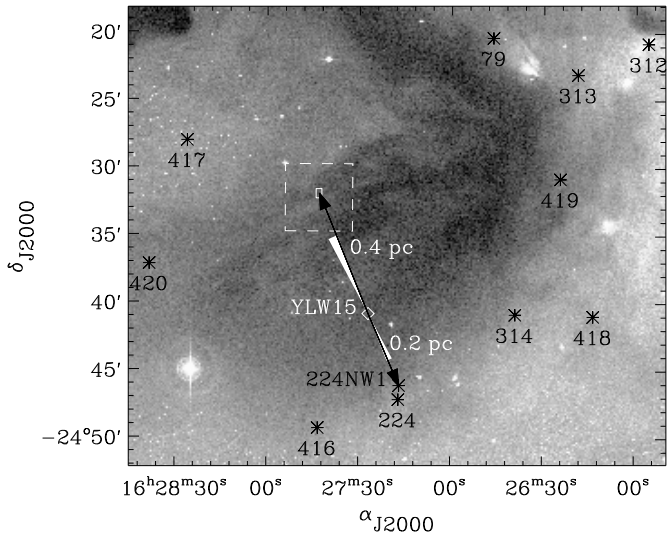}
      \caption{{\footnotesize Optical finding chart (DSS2-red) of the HH objects in the 
$\rho$ Ophiuchi dark cloud. The numbers refer to the HH object list of Reipurth (1999).
The dashed square shows the NTT/SOFI field of view.
The position of the new embedded HH objects is marked by the white box.
The white sectors show the position angle of the thermal jet candidate of the Class~I 
protostar YLW15 including 90$\%$ error level (Girart et al.\ 2000).}
}
         \label{optical_HH}
   \end{figure}

Fig.~\ref{hh}.c shows that the Class~I protostar \object{YLW15}, 
also called \object{IRS43}, is on-axis of the knots A--B, 
$\sim$10$\arcmin$ away (0.4\,pc for d=140\,pc).
This evolved protostar is a radio source, and  was recently announced to be a 
subarcsecond radio binary (\cite{girart00}). 
The main radio component, \object{YLW15-VLA1}, is spatially extended with a position 
angle of 25\degr$\pm$2\degr (one sigma error), and Girart et al. proposed it 
to be a thermal radio jet candidate.
The position angle (PA) of the knot chain, 22.3\degr, is compatible with the PA 
of this jet candidate within errors (Fig.~\ref{optical_HH}).
As the probability for a coincidence by chance between these PAs 
is very low (2$\times$1.645$\times$2\degr/180\degr$\approx$0.037), 
the identification of YLW15 as the exciting source of knots A--B is likely. 
Fig.~\ref{optical_HH} displays the position of the HH objects of this dark cloud.
The [S\,{\small II}] knot \object{HH224NW1} (\cite{gomez98}) was associated with 
the complex HH object \object{HH224}. We note however that HH224NW1 is displaced 
from the axis of HH224, and that 
the PA of YLW15 is 23.4\degr, thus also compatible with the orientation of 
the thermal jet candidate. HH224NW1 may then be related to a counterjet of YLW15.
This association between YLW15 and these HH objects strengthens the proposal of 
Girart et al. for a thermal jet in this object. 
Moreover the alignment with the HH objects moving at $V\le$10--20\,km\,s$^{-1}$ 
implies that this jet is not precessing at least on timescale 
$\sim$(2--4)$\times$10$^4$\,yr.

These observations confirm that HH objects excited by Class I protostars may be hidden 
by large extinctions in the optical, but can be easily unveiled and studied by 
deep near-IR observations. Embedded HH objects are probably very common, 
and it is essential now to obtain a reliable census, for 
instance to quantify the role of outflows in maintaining a high level of 
turbulence in molecular clouds and regulating star formation (\cite{matzner00}).

\begin{acknowledgements}
We would like to thank the referee P.\,T.\,P. Ho for his useful comments, 
M.\,D. Smith who provided us the shocked-H$_2$ line fluxes 
partly published in \cite*{smith95}, and the NTT-team for its efficient support 
during the observations.
N.\ G.\ is supported by the European Union (HPMF-CT-1999-00228). 
R.\ N.\ acknowledges financial support from the BMBF through DLR grant 50 OR 0003.
\end{acknowledgements}

\end{document}

%% file: definition_journaux_aasms4.tex
\def\aj{AJ}			
		
\def\apj{ApJ}			
\def\apjl{ApJ}

\def\aap{A\&A}

\def\mnras{MNRAS}

\def\pasp{PASP}